\documentclass[10pt]{article}
\usepackage{hyperref}
 \usepackage{footnote}
\makesavenoteenv{minipage}

\title{Gravitational Collapse in Gravity's Rainbow}
\author{Ahmed Farag Ali$^{1,2}$, Mir Faizal$^{3}$, Barun Majumder$^{4,5}$, Ravi Mistry$^{6}$ \\ \\
 $^1$Department of Physics, Florida State University, \\ Tallahassee, FL 32306, USA\\
$^2$Deptartment of Physics, Faculty of Science, \\ Benha University,
Benha, 13518, Egypt\\
$^3$Department of Physics and Astronomy, University of Waterloo, \\  Waterloo,
Ontario, N2L 3G1, Canada\\
$^4$Department of Physics, IIT Gandhinagar, \\ Ahmedabad, 382424, India \\
$^5$Deptartment of Physics, Montana State University, \\ Bozeman, MT  59717,  USA\\
$^6$Department of Physics and Electronics, St. Xavier's College, \\ Ahmedabad, 380009, India}
\date{}

\begin{document}

\maketitle

\begin{abstract}
In this paper, we will analyze the gravitational collapse in the framework of      gravity's rainbow.
We will demonstrate  that the position of the horizon for a particle inside the black hole
depends on the energy of that particle. It will also be   observe that the position of the horizon for a
particle   falling radially  into the black hole   also depends on its energy.
Thus, it is possible for a particle coming from outside to interact with a particle   inside the black, and take
some information  outside the black hole. This is because for both these particles the
position of horizon is different. So, even though the particle from inside the black hole is in its own horizon,
it is not in the horizon of the particle coming from outside. Thus, we will demonstrate that in gravity's rainbow
information can get out of a black hole.
\end{abstract}

\maketitle

The formation of a  black hole occurs when a    massive star runs out of nuclear fuel and
   collapses because of its own gravitational force \cite{g1}-\cite{g2}.
 The gravitational collapse causes the formation of an event horizon, and this prevents
 the information from getting out of the black hole.
 However, in course of time the black hole is expected to evaporate as it emits  a thermal radiation
 called the Hawking radiation.
 Thus, we arrive at the
information paradox  in black holes  \cite{1}-\cite{q}.
In this paper, we will analyse the  black hole information paradox of black holes
in the context of gravity's rainbow.
We will demonstrate that it
is possible for information to get out of a black hole in gravity's rainbow.
It is expected from almost all approaches to quantum gravity that spacetime has a minimum measurable length
\cite{d}-\cite{d1}. This minimum length scale is usually taken to be the Planck length $L_p$
The limit on the scale at which spacetime can be measured can be translated into a limit
on the   energy of  an particle that is being used to probe this spacetime. This is because
if the energy $E $ is required to probe spacetime at a length scale $L $, then we require an energy greater
than $E $ to probe spacetime at length scale less than $L$. Now if spacetime has a minimum measurable length
scale $L_p$, and the energy used to probe spacetime at this length scale is $E_p$, then we cannot have
an energy more than $E_p$.  The existence of this maximum energy
can be incorporated into a deformed version of special
relativity called   the doubly special relativity theory
 \cite{2}-\cite{3}.
 It is also possible to generalize this doubly special relativity to
 curved manifolds and the resultant theory is called
gravity's rainbow \cite{n1}-\cite{n2}.
It may be noted that various black hole solutions have been studied in the framework of
gravity's rainbow \cite{m1}-\cite{m2}. It has been demonstrated that the  black hole thermodynamics gets
modified in gravity's rainbow.

In gravity rainbow the energy-momentum dispersion relation is modified as
\begin{equation}
 E^2 f(E/E_{p})^2- p^2 g(E/E_{p})^2= m^2,
\end{equation}
where $E_{p}$ is the Planck energy scale, $m$ is the mass of the test particle. Here the energy dependent functions
 $f(E/E_{p})$ and $g(E/E_{p})$,
 are the rainbow functions, and they are required to satisfy
\begin{eqnarray}
 \lim_{E\rightarrow 0} f(E/E_P) =1, && \lim_{E\rightarrow 0} g(E/E_P) =1.
\end{eqnarray}
Now the energy dependent contribution in the  metric
can be incorporated by defining
\begin{equation}
h(E/E_P) = \eta^{ab} e_a(E/E_{p}) \otimes e_b(E/E_{p}),
\end{equation}
where $e_0(E/E_{p}) =(1/f(E/E_{p})) \tilde{e}_0$ and $e_i(E/E_{p})= (1/f(E/E_{p})) \tilde{e}_i$.

There are  various choices  of  phenomenologically motivated  Rainbow functions
\cite{allref}-\cite{temp}.   The
hard spectra from gamma-ray burster's at cosmological distances has been used to motivate a certain type of
  rainbow functions  \cite{amea}
  \begin{eqnarray}
    f(E) = (e^{\alpha E/E_p } -1)   \left(\alpha  E/E_p\right)^{-1}, && g(E) = 1.
  \end{eqnarray}
It is also possible to choose rainbow functions where the   velocity of light is constant are given by
\cite{2}
\begin{eqnarray}
   f(E)  ( 1 - \lambda E/E_p)^{-1}, &&  g(E) = ( 1 - \lambda E/E_p)^{-1}.
\end{eqnarray}
It is also possible to use results from loop quantum gravity and non-commutative geometry to motivate
another type of rainbow functions \cite{ame}-\cite{ame1}
\begin{eqnarray}
 f(E)=1, && g(E)=\sqrt{1-\eta(E/E_P)^n}.
\end{eqnarray}
 However, the main property of all these rainbow functions is that they make
 spacetime energy dependent and this is what we require for our analysis.

In this paper, we study the gravitational collapse in the framework of Gravity's Rainbow.
So, we will analyse the  modified Einstein field equations for the gravitational field in a given
material. By taking the medium to be dust-like, then we may neglect the pressure. The spherically symmetric line
element in the framework of gravity's rainbow is given as
\begin{eqnarray}
ds^2=&-& \frac{c^2 d\tau^2}{f^2(E)}+ \frac{e^{\lambda(\tau,R)} dR^2}{g^2(E)}
+\frac{r^2(\tau,R) d\theta^2}{g^2(E)}+ \frac{r^2(\tau,R) \sin^2\theta d\phi^2}{g^2(E)}.
\end{eqnarray}
The function $r(\tau,R)$ is the radius metric. It is noted that the metric above fixes the choice of $\tau$ and allows for arbitrary choice of
the radial coordinates $R\to R^{\prime}$.
From the considered metric, it has been calculated the independent components of Einstein Field equation as follows\footnote{We have used c=1 unit system.} :
\begin{eqnarray}
\frac{g^2(E)}{f^2(E)}\left[\frac{1}{r^2}+e^{-\lambda}\left(\frac{{\lambda}^\prime {r}^\prime}{r}
-\frac{2 r^{\prime\prime}}{r}-\frac{{r^{\prime}}^2}{r^2}\right)\right]+\frac{\dot{r}\dot{\lambda}}{r}
+\frac{\dot{r}^2}{r^2}&=&8 \pi G \varepsilon, ~~~~~\label{1}\\
e^{-\lambda}{r^{\prime}}^2- \frac{f^2(E)}{g^2(E)}\left(2 \ddot{r} r+\dot{r}^2\right)-1&=&0,  \label{2}\\
\frac{e^{-\lambda}}{r} (2 r^{\prime\prime}-\lambda^{\prime} r^{\prime})-\frac{f^2(E)}{g^2(E)}\left(\frac{2 \ddot{r}}{r}+\ddot{\lambda}+\frac{\dot{\lambda}^2}{2}+\frac{\dot{r}\dot{\lambda}}{r}\right)&=&0~~~\label{3}~~~\\
2 \dot{r}^\prime-\dot{\lambda} r^{\prime}&=&0, \label{dotr}
\end{eqnarray}
where the prime represents derivative with respect to $R$, and the dot with respect to $\tau$.

We will integrate  Eq. (\ref{dotr}) with respect to time $\tau$, and thus obtain,
\begin{eqnarray}
e^{\lambda}=\frac{{r^{\prime}}^2}{1+a(R)}, \label{tau}
\end{eqnarray}
where $a(R)$ is an arbitrary function such that $1+a(R)> 0$.
By substituting Eq. (\ref{tau}) into Eq. (\ref{2}), we get
\begin{eqnarray}
a-2 \ddot{r} r \frac{f^2 (E)}{g^2 (E) }- \dot{r}^2 \frac{f^2 (E)}{g^2 (E)}=0.
\end{eqnarray}
By integrating the last equation, we get:
\begin{eqnarray}
\dot{r}^2= \frac{a(R)g^2 (E)}{f^2 (E)}+\frac{A(R)}{r},
\end{eqnarray}
where $A(R)$ is another arbitrary function.
Now from the above equation we can get the value of $r$,
\begin{eqnarray}
r&=&\frac{A f^2 (E) }{2 a g^2 (E)  } \left[\cosh\eta -1 \right],~~ \rm{for}~~~ a > 0, \\
r&=&\frac{A f^2 (E) }{-2 a g^2 (E) } \left[1-\cos\eta \right],~~ \rm{for}~~~ a < 0, \\
r&=&\frac{(9A)^{1/3}}{(4)^{1/3}}\left[\tau_0 (R) -\tau\right]^{2/3}, ~~\rm{for}~~ a=0.
\end{eqnarray}
Furthermore, the time is given by
\begin{eqnarray}
\tau= \pm \int \frac{dr}{\sqrt{\frac{g^2(E)a(R)} {f^2 (E) }+\frac{A(R)}{r}}}.
\end{eqnarray}
So, we can write
\begin{eqnarray}
\tau_0 (R)-\tau&=& \frac{A f^3 (E)}{2a^{3/2} g^3 (E) }\left[\sinh\eta - \eta\right],~~ \rm{for}~~a > 0, ~~~~~\\
\tau_0 (R)-\tau&=& \frac{A f^3 (E) }{2 (-a)^{3/2} g^3 (E) }\left[ \eta - \sin\eta\right],~\rm{for}~a < 0, ~~~~\\
\tau_0 (R)&=&2\tau + C_1,~~~~~~~~~~~~~~~~\rm{for}~~~~a = 0.
\end{eqnarray}
Where $C_1$ is a finite integration constant. The important point in this analysis is that the position of the horizon depends on the energy of particle.
So, particles of different energies will see the horizon form differently in gravity's rainbow.

Now a similar situation occurs for a particle outside the black hole. Thus, for
an   particle falling radially  into the black hole  with velocity
$v^{\mu}= dx^{\mu}/{ds}$, we can write
\begin{equation}
\frac{dv^0}{ds} = - \Gamma^0_{\mu \nu}(E)~v^{\mu}v^{\nu}.
\end{equation}
Here we have assumed   $v^2 = v^3 =0$, as the particle is falling radially  into the black hole.
The time taken by the particle to reach the black hole can be calculated
from this expression \cite{epl}
\begin{equation}
t = -\frac{2M f(E)}{g(E)} \log(r-2 G M) - \frac{(2k^2 f^2(E)+1)r}{2f(E)g(E)k^2} + C_2, 
\end{equation}
Where $C_2$ is a finite integration constant. Thus, the position of the horizon for a particle outside the
black hole also depends on the energy of the particle.

Now the position of the
horizon for the particle inside and outside the black hole depends on the energy of those particles.
Thus, if there are two particles, say $P_1$ and $P_2$, such that $P_1$ is inside the black hole and $P_2$
is outside the black hole. Now the position of the horizon for both $P_1$ and $P_2$ depends on the energy
of $P_1$ and $P_2$. Let us assume that the particle $P_1$ is inside its horizon, which has formed at a
distance  $r_1$. However, it is now possible that the particle $P_2$ has a different value for the horizon
which is at a distance $r_2$, such that $r_2 < r_1$. In this case, it is possible for the particle
$P_2$ to come close to the black hole and interact with the particle $P_1$, and leave with information
contained in $P_1$. This is because even though the particle $P_1$ is inside its horizon, the particle
$P_2$ is not inside its horizon. Thus, information can be carried out of the black hole by the particle
$P_2$, as the position of the horizon is an energy dependent concept in gravity's rainbow.
  So, in the   gravity's rainbow, information can go out
even after the collapse.


\begin{thebibliography}{99}
 \bibitem{g1}J. R. Oppenheimer and H. Snyder, Phys. Rev. {\bf 56} 55 (1939).
 \bibitem{gh} C.  Vaz Nucl.  Phy. B {\bf  891}  558 (2015).
 \bibitem{hg} G. Pinheiro and  R. Chan,   Gen. Rel. Grav.  {\bf43} 145 (2011).
 \bibitem{g2} P. C. Vaidya  Phys. Rev. {\bf 83} 10 (1951).
\bibitem{1} S. W. Hawking, Phys. Rev. D {\bf 14} 2460 (1976).
\bibitem{a} L. Susskind, L. Thorlacius and J. Uglum, Phys. Rev. D {\bf 48} 3743 (1993).
\bibitem{b} X. Calmet,   Class. Quantum Grav. {\bf 32} 045007 (2015).
\bibitem{b1}M.  Faizal,  Int. J. Geom. Meth. Mod. Phys. {\bf 11} 1450010  (2014).
\bibitem{a1}A. Sen,   JHEP {\bf 1011}  138 (2010).
\bibitem{q} K. Papadodimas and  S.  Raju,  Phys. Rev. Lett. {\bf 112}  051301 (2014).

\bibitem{d} L. J. Garay, Int. J. Mod. Phys. A {\bf 10} 145 (1995).
\bibitem{d1} A. Ashoorioon, A. Kempf and R. B. Mann, Phys. Rev. D {\bf 71} 023503 (2005).
\bibitem{2} J. Magueijo and L. Smolin, Phys. Rev. Lett. {\bf 88} 190403 (2002).
\bibitem{21} J. Magueijo and L. Smolin, Phys. Rev. D  {\bf 71} 026010 (2005).
\bibitem{3} J. L. Cortes and J. Gamboa, Phys. Rev. D {\bf 71} 065015 (2005).
\bibitem{n1} J. Magueijo  and L. Smolin, Class. Quant. Grav. {\bf 21} 1725 (2004).
\bibitem{n2} J. J. Peng and S. Q. Wu, Gen. Rel. Grav. {\bf 40} 2619 (2008).
\bibitem{m1}A. F. Ali, M. Faizal and M. Khalil,  JHEP {\bf 1412}  159  (2014).
\bibitem{m1ab}A. F. Ali, M. Faizal and M. Khalil, Nucl. Phys. B {\bf 894} 341 (2015).
\bibitem{m1cd}A. F. Ali, Phys. Rev. D {\bf 89}  104040  (2014).
 \bibitem{m1ef}  Y. Gim and  W.  Kim,  JCAP {\bf 10} 003 (2014).
\bibitem{m2} A. F. Ali, M. Faizal and M. Khalil, Phys. Lett. B {\bf 743} 295 (2015).

\bibitem{allref} A. Awad, A. F. Ali and B. Majumder, JCAP {\bf 10} 052 (2013).
\bibitem{allref1} R. Garattini and B. Majumder, Nuclear Physics B {\bf 883} 598 (2014).
\bibitem{allref2}B. Majumder, Int. J. Mod. Phys. D {\bf 22} 1342021 (2013) .
\bibitem{allref3}G. Amelino-Camelia, M. Arzano, G. Gubitosi and J. Magueijo, Phys. Rev. D {\bf 88} 041303 (2013).
\bibitem{allref4}J. D. Barrow and J. Magueijo, Phys. Rev. D {\bf 88} 103525 (2013).
\bibitem{allref5}R. Garattini and B. Majumder, Nuclear Physics B {\bf 884} 125 (2014).
\bibitem{allref6}C. Leiva, J. Saavedra and J. Villanueva, Mod. Phys. Lett. A {\bf 24} 1443 (2009).
\bibitem{allref7}R. Garattini and G. Mandanici, Phys. Rev. D {\bf 85} 023507 (2012) .
\bibitem{allref8}R. Garattini and G. Mandanici, Phys. Rev. D {\bf 83} 084021 (2011)   .
\bibitem{temp} A. F. Ali, Phys. Rev. D {\bf 89} 104040 (2014).

\bibitem{amea}G. Amelino-Camelia, J. R. Ellis, N. Mavromatos, D. V.
Nanopoulos  and S. Sarkar, Nature {\bf 393} 763 (1998).
\bibitem{ame} G. Amelino-Camelia, Living Reviews in Relativity {\bf 16}, 5 (2013).
\bibitem{ame1} U. Jacob, F. Mercati, G. Amelino-Camelia and T. Piran, Phys. Rev. D {\bf 82} 084021 (2010).
 \bibitem{epl} A. F. Ali, M. Faizal and B. Majumder,  Europhys. Lett. {\bf 109} 20001 (2015).
 \end{thebibliography}
\end{document}